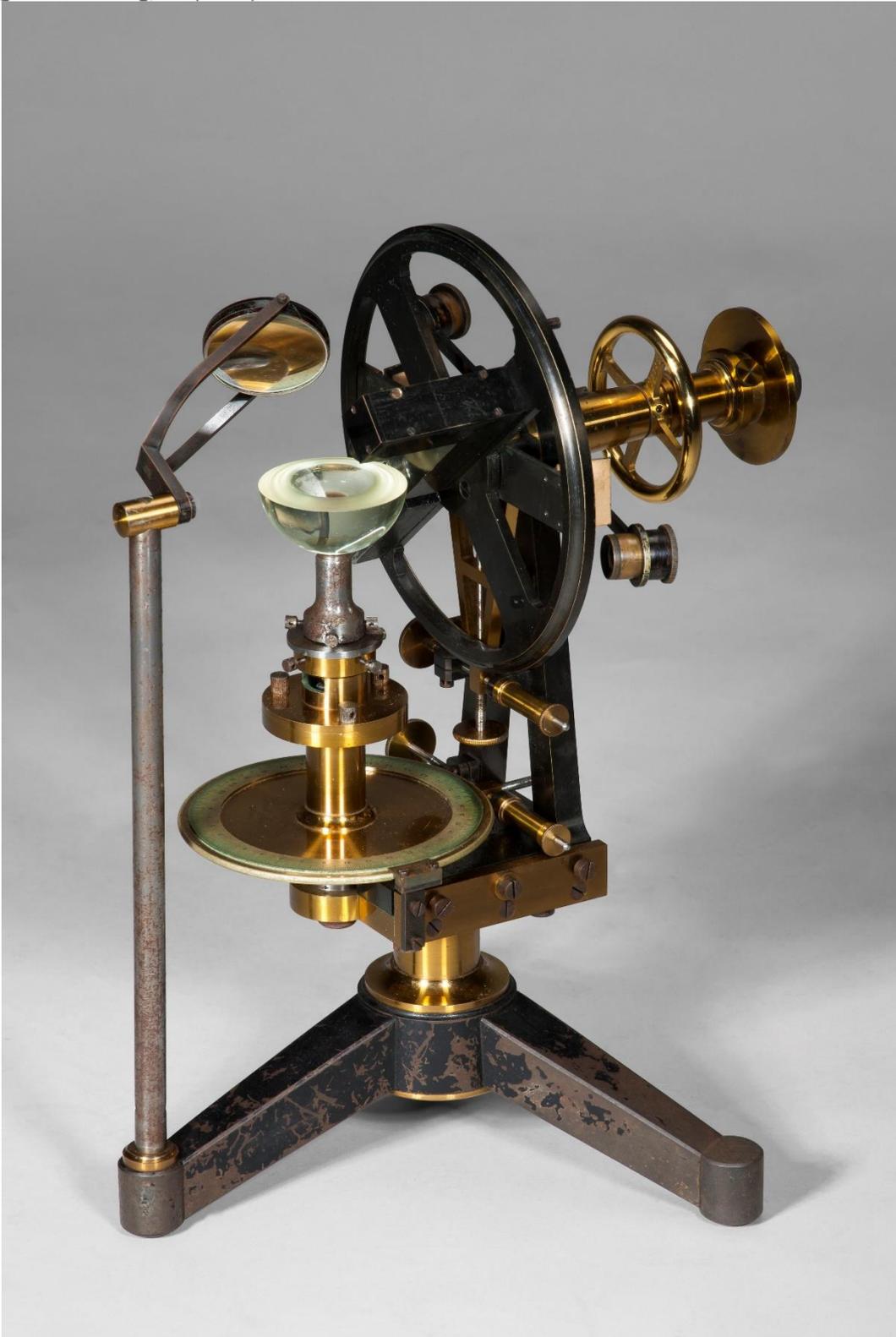

# Physikalische Sammlung

Als Gerätesammlung diente die Physikalische Sammlung ursprünglich zur Demonstration physikalischer Phänomene während der Ausbildung im Fach Physik; ihre Anschaffung steht in enger Verbindung mit der Einführung des Lehrfaches Experimentalphysik an der Philipps-Universität Marburg.

Schon kurz nach Gründung der Philipps-Universität im 16. Jahrhundert wurde eine Professur für das Fach Physik eingerichtet. Die Physik verstand man gemäß Schmitz noch ganz im alten aristotelischen Sinne als „die Lehre von allen und jeden naturen sowol Göttlichem und auch geistigen, als auch cörperlichen" und ordnete sie als ein reines Lehrfach der Philosophischen Fakultät zu. Bis weit in das 17. Jahrhundert hinein beschränkte sich die Praxis des Experimentierens vornehmlich auf die Astronomie und nur wenige Universitäten boten dies an. In die Zeit des ausgehenden 17. Jahrhunderts fiel jedoch eine bemerkenswerte Änderung in der Verfügbarkeit von Geräten zur Veranschaulichung naturwissenschaftlicher Zusammenhänge und nach und nach fand das Experimentieren an den Europäischen Universitäten Zuspruch.

Von Jacob Waldschmied (1644–1689), der als Primarius der Medizin 1682 auch das Fach Physik übernahm, wird bei Schmitz als erstem berichtet, er habe in seinen Vorlesungen Vorführungen „der Phänomene der Luft" gezeigt. Leider sind keine Aufzeichnungen über die dabei verwendeten Geräte überliefert. Der Mediziner Johann Dorstenius (1643–1706), der ab 1690 ad interim die Physikprofessur innehatte, trat als Nachfolger Waldschmieds 1695 die ordentliche Physikprofessur an. Zu dieser Zeit baute im niederländischen Leiden Professor Burchard de Volder (1643–1709) das erste europäische „theatrum physicum" mit einer Gerätesammlung aus der Werkstatt des dort ansässigen

**gegründet 1694**
**Sammlung** Historische Geräte zur Verwendung in Lehre und Forschung im Fach Physik, vornehmlich aus dem 19. und 20. Jahrhundert. Ältestes Gerät ist eine Pumpe, die in Teilen vermutlich aus den Anfängen der Sammlung (1694) stammt
**Anzahl der Objekte** ca. 1000

Linke Seite: Kristall-Refraktometer der Fa. Carl Zeiss Jena, 1902. Das Refraktometer dient zur Untersuchung des Brechungsindexes von Kristallen. Im Katalog der Fa. Carl Zeiss Jena von 1893 heißt es: „Die Brechungsindices der zu untersuchenden Substanzen werden aus dem Grenzwinkel der Totalreflexion in einem schweren Flintglas gegen die betreffende Substanz ermittelt." .

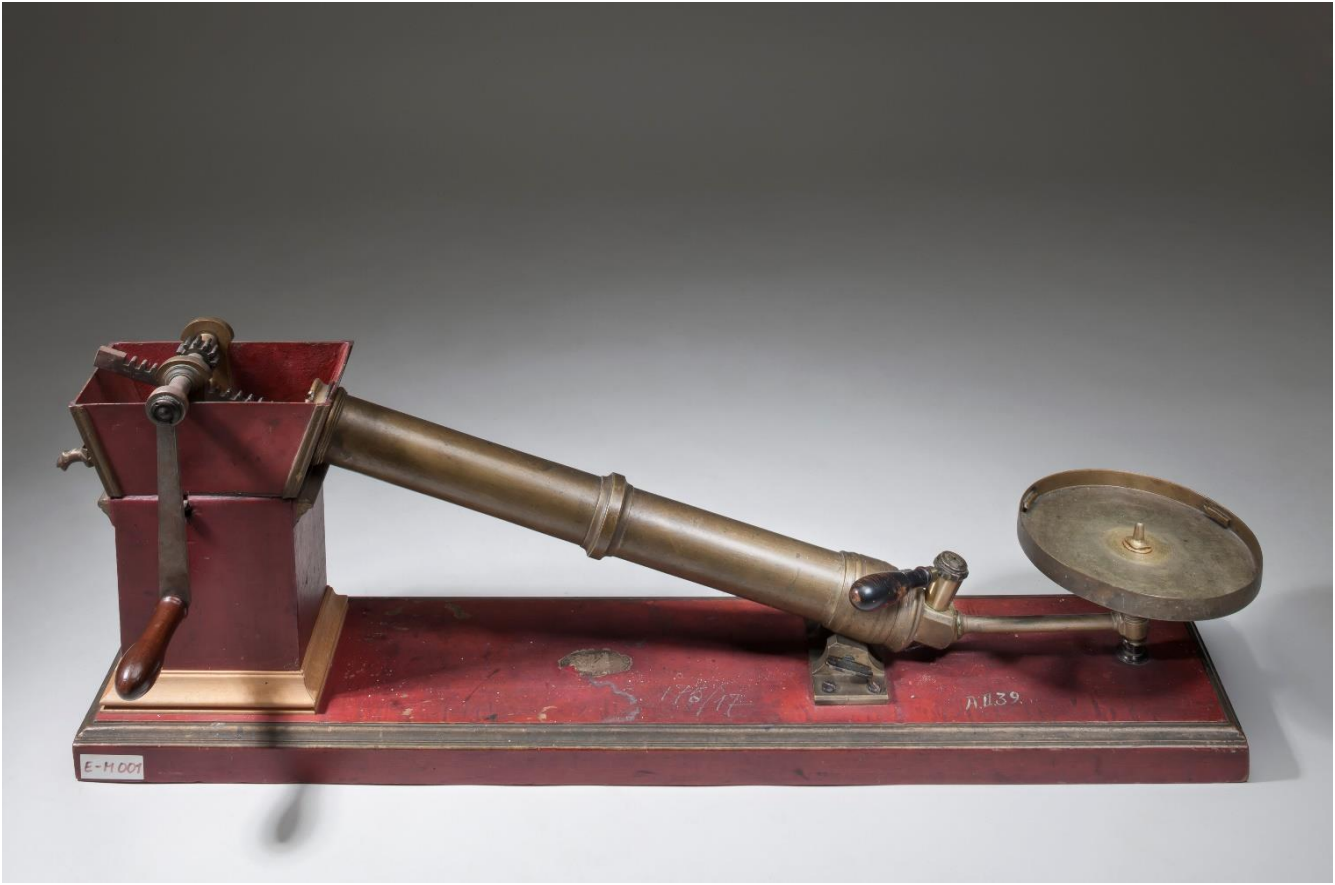

Die Pumpe der Physikalischen Sammlung. Vermutlich ist dies in Teilen die originale Pumpe Musschenbroeks von 1694, die von Johann Gottlieb Stegmann Ende des 18. Jahrhunderts umgebaut und wieder nutzbar gemacht wurde. Die Pumpe ist damit eines der ältesten Geräte der Philipps-Universität.

Johan van Musschenbroek (1660–1707) auf, um den Physik-Studenten „die Wahrheit und Gewissheit der Postulate und Theorien" zu demonstrieren. Leiden avancierte mit den Worten de Clercqs zur „Wiege und Kinderkrippe der Experimentalphysik in Europa". Die Instrumentenwerkstatt der Musschenbroeks entwickelte sich Ende des 17. und Anfang des 18. Jahrhunderts zu einem der größten europäischen Exporteure wissenschaftlicher Geräte. Van Musschenbroek pflegte verschiedene Kontakte nach Deutschland. Er stellte die Geräte in handgeschriebenen Katalogen mit Festpreisen vor, ein Novum im damaligen Handel. So muss auch Dorstenius von dem Instrumentenbauer in Leiden gehört haben. 1693 bat er um ein Angebot für eine Luftpumpe zur Demonstration von Phänomenen der Pneumatik. Von 1694 bis 1696 erfolgten fünf Lieferungen von Geräten, eine weitere im Jahre 1703. Es handelte sich dabei um Geräte zur Pneumatik, Mechanik, Hydrostatik, Optik sowie weitere physikalische Instrumente. Bei der ersten Lieferung befand sich auch die kleinere der beiden Varianten der Luftpumpe aus Musschenbroeks Katalog.

Zwei weitere Instrumentenbauer gehörten zu Dorstenius' Lieferanten: Bei Ehrenfried Walther von Tschirnhaus erwarb er einen Brennspiegel, mit dem er Metalle zum Schmelzen bringen konnte, und von Christian Schober aus Leipzig kaufte er einen Satz mathematischer Instrumente. Aus den Akten geht nicht hervor, dass Dorstenius die Geräte im Auftrag der Universität kaufte; es ist vielmehr ist anzunehmen, dass er alle Lieferungen aus seinen eigenen Mitteln bezahlte. Im Vorlesungsverzeichnis von 1697 kündigte Dorstenius laut Schmitz den neuen Kursus, das „Experimentier-Collegium", an:

„Diesen Weg [das Experimentieren] werde ich jetzt befolgen und dabei den Mantel des Geheimnisses vor vielen Wunderdingen lüften, um damit Gott in der Natur bewundern zu können und Dinge mit Hilfe kostbarer Maschinen, wie sie in Deutschland bislang noch nicht zu sehen waren, vorführen und so ein vollständiges privates Experimentier-Collegium ankündigen."

Johann Dorstenius gilt damit als Begründer der Physikalischen Sammlung und der Experimentalphysik in Marburg. Von den Geräten der Erstausstattung der Sammlung sind vermutlich noch Teile der Pumpe in deren Überarbeitung im Jahre 1785 durch Johann Gottlieb Stegmann erhalten.

Bis in das frühe 19. Jahrhundert hinein wechselte die Besetzung des Lehrstuhls für Physik als Zweitfach mehrmals zwischen Lehrstühlen der Medizin, Mathematik und Philosophie. Dies hinterließ auch Spuren in der Physikalischen Sammlung: Es mussten bei jedem Wechsel Übergaben arrangiert werden, in deren Folge es auch zu gegenseitigen Vorwürfen über mangelnde Pflege und zu Verlusten in der Sammlung kam. Eine größere Erweiterung erlebte die Sammlung 1810 durch die Berufung von Georg Wilhelm Muncke (1772–1847) aus Hannover als Professor für Mathematik und Physik nach Marburg. Durch Auflösung des Georgianums in Hannover und der Universität Rinteln Anfang des 19. Jahrhunderts wurden deren Geräte in die Marburger Sammlung übernommen.

Im Jahre 1817 erfolgte schließlich mit der Berufung von Christian Ludwig Gerling (1788–1864) die Gründung des Mathematisch-Physikalischen Instituts der Philipps-Universität und somit die institutionelle Anbindung der Sammlung. Das Institut war jedoch zunächst in zu kleinen und unzulänglichen Räumen im Deutschen Haus untergebracht. Nach langwierigem Ringen erhielt Gerling 1838 endlich die Zusage, neue und vor allem größere Räumlichkeiten im Hauptgebäude des ehemaligen Dörnberger Hofes am Renthof beziehen zu dürfen. Dort ließ er nicht nur die Geräte der Physikalischen Sammlung thematisch geordnet in einzelnen Zimmern fest und dauerhaft aufstellen, sondern baute zudem den oberen Teil des alten Turms zur Sternwarte der Philipps-Universität um.

Mit einem eigenen Institut wandelte sich auch der Anspruch an das Fachgebiet: Es war nun nicht mehr nur ein Lehrfach, sondern entwickelte sich rasch zu einem Forschungsgebiet. Während Gerlings Schwerpunkte auf der Entwicklung höchst präziser Messmethoden und Auswertungen für geodätische

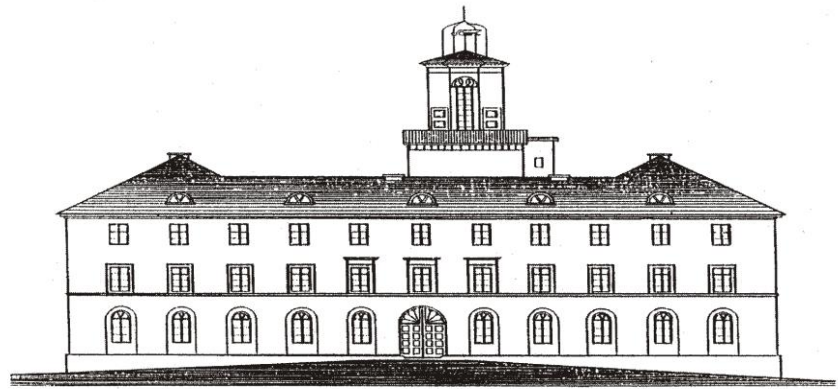

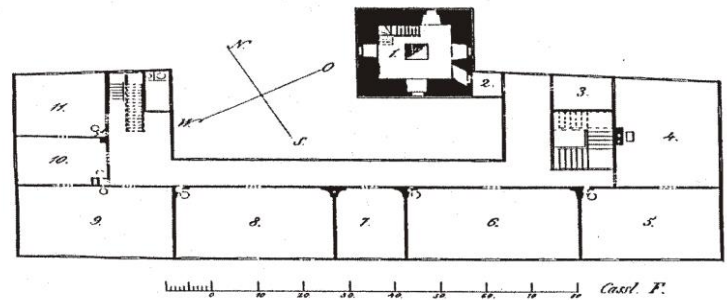

RENTHOF 6
MIT GRUNDRISS I. OBERGESCHOSS
1841

Das neue Mathematisch-Physikalische Institutsgebäude nach Vollendung des Umbaus 1841. Das Gebäude des ehemaligen Dörnberger Hofes liegt am Renthof 6. In seiner Rektoratsrede von 1848 erläutert Gerling die Nutzung der Räume des Instituts, vor allem die Aufstellung der Geräte der Physikalischen Sammlung.

und astronomische Positionsmessungen lagen, kamen schon bald weitere Gebiete hinzu und damit auch Geräte, die nicht in der Lehre sondern in der Forschung eingesetzt wurden. Auch bei den Instrumenten für die Lehre kam es zu einer Veränderung – der Aufteilung in Geräte für die praktischen Übungen der Studierenden während ihrer Ausbildung (also dem frühen Sinn der Sammlung entsprechend) und einer besonderen Gerätesammlung für Vorführexperimente in den Vorlesungen. Die beiden letztgenannten Geräteparks werden heute nicht mehr zur Physikalischen Sammlung gezählt.

Die Physikalische Sammlung umfasst heute die noch vorhandenen aber nicht mehr genutzten älteren Geräte zur Ausbildung der Studierenden sowie einige der frühen, zu Forschungszwecken eingesetzten Messinstrumente. Zu den besonderen Schätzen unter den frühen Geräten zählen:

– eine mechanische Pumpe, in Teilen vermutlich aus dem Jahre 1694

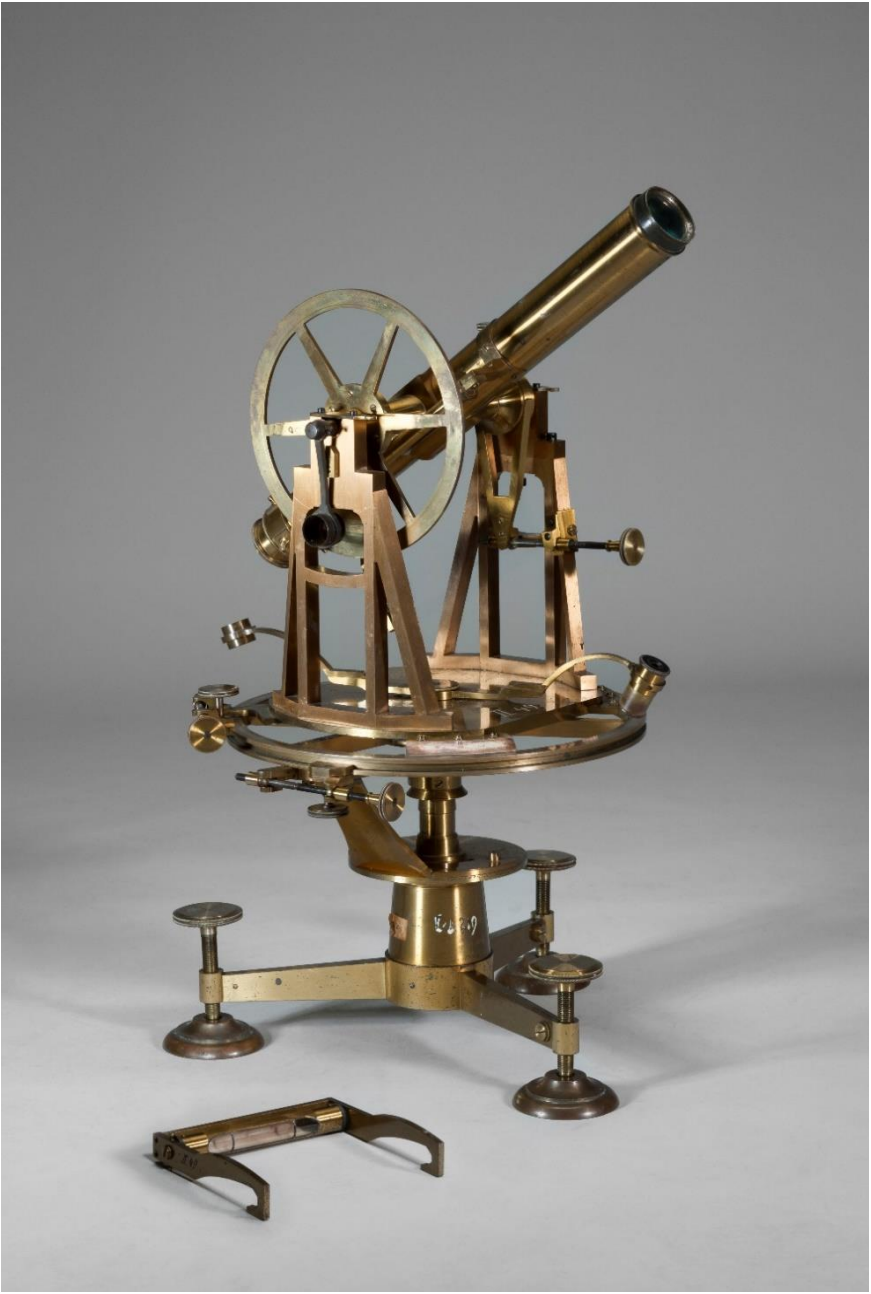

Universalinstrument oder auch Repetitionstheodolit von 1823, hergestellt von der Firma Breithaupt in enger Kooperation mit Christian Ludwig Gerling; eines der Messgeräte der Haupttriangulation Kurhessens Anfang des 19. Jahrhunderts.

- ein Theodolit der Firma Breithaupt, 1823, eingesetzt während der Kurhessischen Triangulierung
- eine Toise du Pérou, Normmaß, 1831, während der Kurhessischen Triangulierung verwendet
- ein Sinus-Elektrometer von Rudolf Kohlrausch, 1857, Teil des berühmten Kohlrausch-Weber-Experiments zur Bestimmung der Lichtgeschwindigkeit

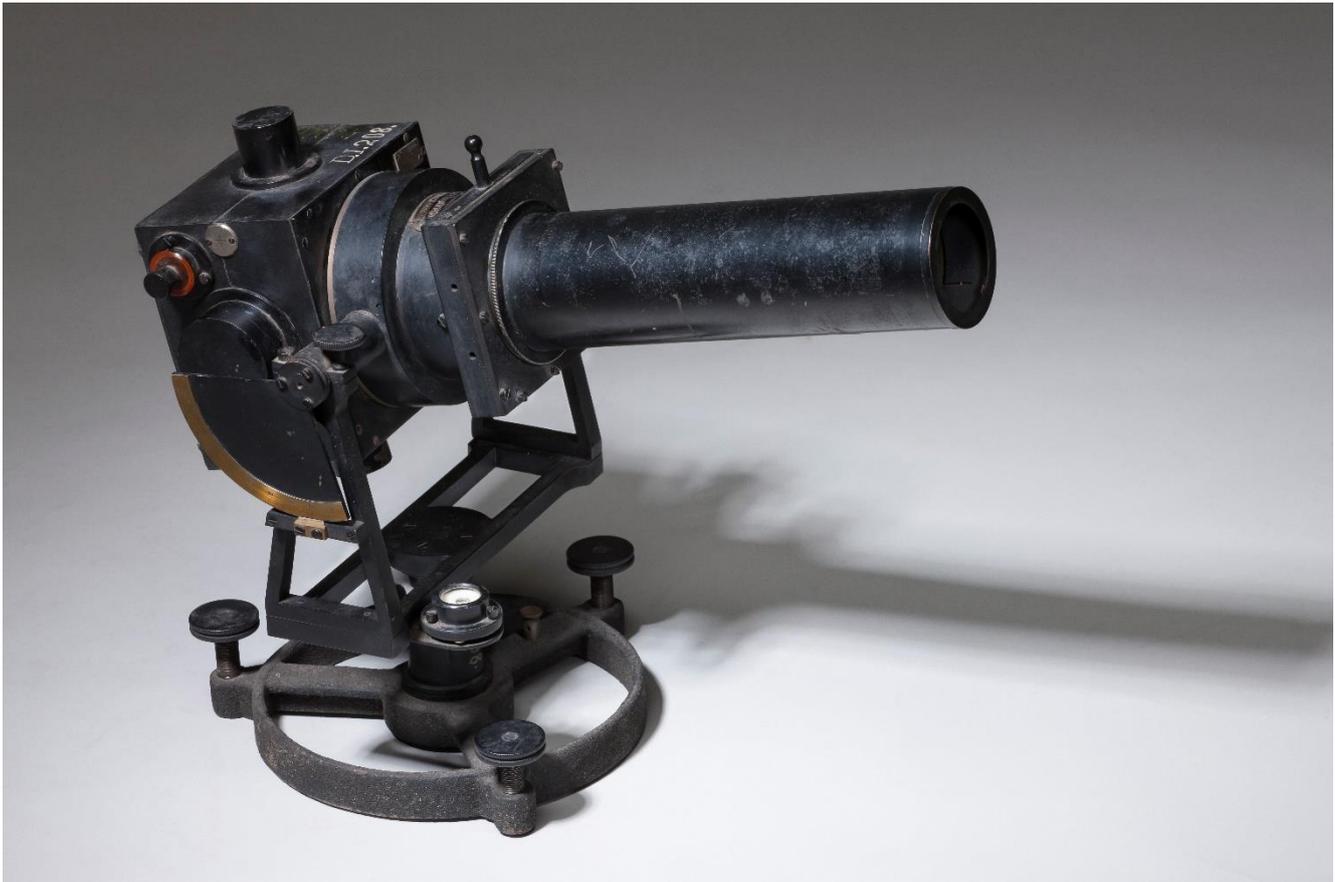

Photometer der Firma Günther & Tegetmeyer, Braunschweig, 1915. Eines der drei noch bekannten vorhandenen Photometer mit den ersten Photozellen nach Elster und Geitel.

– ein Inklinatorium von Moritz Meyerstein, Göttingen, 1858, ab Mitte des 19. Jahrhunderts eingesetzt zur Messung der Inklination des Erdmagnetfeldes, im Zuge einer Kooperation mit dem Magnetischen Verein
– ein Kristall-Refraktometer, Carl Zeiss, Jena, 1902
– ein Photometer der Firma Günther & Tegetmeyer, Braunschweig, 1915; eines der drei noch bekannten vorhandenen Photometer mit den ersten Photozellen nach Elster und Geitel.

Eine kleine Dauerausstellung der Sammlung befindet sich im Renthof 6, im Erdgeschoss des Turms der Sternwarte. Insgesamt umfasst die Sammlung etwa 1000 Geräte zu verschiedenen Gebieten der Physik.

*Andreas Schrimpf*

## Literatur

## Physical Collection

At the end of the 17th century, the collection was established through the acquisition of Leiden instrumentmaker Musschenbroek's collection. Its aim was to demonstrate physical phenomena.

Until the end of the 18th century, the collection was used and extended by a number of professors from various university departments in teaching mechanics, hydrostatics, optics and electrostatics. The foundation of the Institute for Mathematics and Physics in 1817 and the appointment of Christian Ludwig Gerling to the position of Professor for Mathematics, Physics and Astronomy strengthened the subject considerably. Gerling extended the collection mainly with devices for geodesy and astronomy, with more instruments being added in the time up to the middle of the 20$^{th}$ century. Today, the collection allows an insight into the history of scientific instruments as witnesses of the past.

**Physikalische Sammlung**
Fachbereich Physik der Philipps-Universität Marburg
Renthof 6, 35032 Marburg
Ansprechpartner: Priv. Doz. Dr. Andreas Schrimpf
Tel: +49 6421 28-21338
E-Mail: andreas.schrimpf@physik.uni-marburg.de
*http://www.uni-marburg.de/sammlungen/sammlungen/physik*

Besichtigung nach Vereinbarung.